\magnification=1200
\hsize =32true pc\vsize = 48true pc
\hoffset=.375 true in
\voffset=.5true in

\font\mysmall=cmr8 at 8 pt
\font\eightit=cmti8

\def\c{\bf C}

\def\z{\bf Z}

\def\r{\bf R}

\def\hh{\cal H}

\def\oo{\cal O}

\def\oo{\cal O}

\def\mm{\cal M}

\def\cc{\cal C}

\def\ab{\cal A}
\def\bb{\cal B}
\def\cc{\cal C}
\def\picture #1 by #2 (#3){
		\vbox to #2{
		\hrule width #1 height 0pt depth 0pt
		\vfill
		\special {picture #3}}}
\font\teneufm eufm10 
\font\seveneufm eufm7 
\font\fiveeufm eufm5
\newfam\eufm
\textfont\eufm\teneufm
\scriptfont\eufm\seveneufm
\scriptscriptfont\eufm\fiveeufm

\def\picture #1 by #2 (#3){
		\vbox to #2{
		\hrule width #1 height 0pt depth 0pt
		\vfill
		\special {picture #3}}}
\centerline
{\bf ON THE CONCEPT OF EPR STATES AND THEIR STRUCTURE}\vskip 0.3 true in
{\hfill{\eightit In  memory of our friend Moshe Flato}\footnote{$^\ast
$}{\mysmall This article is  dedicated to the memory of Moshe
Flato who passed away on November 27, 1998. His deep scientific culture and
unbounded generosity of spirit made a huge impact on many mathematicians and
physicists, including ourselves. We mourn his loss and offer this small
contribution to commemorate his invincible spirit.} \vskip 0.2 true in
\centerline {\bf Richard Arens} \centerline {\mysmall Department of
Mathematics, University of California,Los Angeles}\smallskip
\centerline {\mysmall 
CA 90095--1555,  USA.} 
\bigskip\centerline {\bf  V. S. Varadarajan}
\centerline
{\mysmall Department of Mathematics, University of California,Los Angeles}
\smallskip
\centerline {\mysmall 
CA 90095--1555,  USA}
\bigskip
\noindent
{\mysmall In this paper the notion of an EPR state for the composite $S$ of
two quantum systems $S_1,S_2$, relative to $S_2$ and a set ${\oo}$ of
bounded observables of $S_2$,  is introduced in the spirit of the classical
examples of Einstein--Podolsky--Rosen and Bohm. We  restrict ourselves mostly
to EPR states of finite norm. The main results are contained in Theorem 
3,4,5,6 
in section {\bf III\/} and imply that  if EPR states  of
finite norm relative to
$(S_2,{\oo})$  exist, then the elements of ${\oo}$ have discrete
probability distributions  and the Von Neumann algebra
generated by ${\oo}$ is essentially imbeddable inside $S_1$ by an antiunitary
map. The
 EPR states  then correspond to the different
imbeddings and certain additional parameters, and are explicitly given by
formulae which generalize the famous example of Bohm. If ${\oo}$ generates
all bounded observables, $S_2$ must be of finite dimension and can be 
imbedded inside $S_1$ by an antiunitary map, and the EPR states relative to
$S_2$ are then in canonical bijection with the different
imbeddings of $S_2$ inside $S_1$; moreover they are then given by formulae
which are exactly those of the generalized Bohm states. The notion of EPR
states of infinite norm is also explored and it is shown that the original
state of Einstein--Podolsky--Rosen can be realized as a renormalized limit of
EPR states of finite quantum systems  considered by Weyl,
Schwinger, and many others. Finally, a family of states of infinite norm
generalizing the Einstein--Podolsky--Rosen example is explicitly given.}
\bigskip
\noindent  {\bf I. Introduction\/} \bigskip  Let $S_1, S_2$ be two
quantum systems, for example, those of two one--dimensional particles. The
famous example, first introduced by Einstein, Podolsky, and Rosen in
1935$^1$, describes a state $\sigma $ of the composite system $S=S_1\times
S_2$ with the following property. Let $X_i, P_i$ be the position and momentum
coordinates of the $i^{\rm th}$ particle $(i=1,2)$; then, if a measurement of
$X_1$ (resp. $P_1$) is known to have a definite value when $S$ is in the state
$\sigma $,  the value of $X_2$ (resp. $P_2$) can be predicted with certainty.
The conclusions that these authors drew from this example about the
completeness of the quantum mechanical description of physical reality, and
their refutation by Bohr in 1935$^2$, are well known, and the reader may refer
to the papers of these authors and other related articles on quantum
measurement theory reprinted in the well known reprint collection of
Wheeler--Zurek$^3$.  \medskip 
The Einstein--Podolsky--Rosen state has infinite norm and so does not lie
in Hilbert space; indeed in their example both systems are  infinite
dimensional  and the state in question is actually a distribution state.  In 
an effort to simplify the discussion of Einstein et al, Bohm introduced  spin 
(or polarization) states of particle pairs with the same properties as their
states. Bohm's  example deals with $2$--dimensional quantum
systems and his computations  of the probabilities and discussions of
gedankenexperiments eventually led to experimental tests whether these
probabilities could be derived from a  local hidden variable theory. For
all this the reader may consult Bohm's famous book$^4$ as well as the nice
discussion in$^5$.  
\medskip
In this paper we introduce the concept of a state $\sigma $ of the composite
$S$ of  two quantum systems $S_i(i=1,2)$ being EPR relative to $(S_2,{\oo})$
where
${\oo}$ is any set of  bounded observables of $S_2$. Briefly, this is the case
if there is, for each $A_2\in {\oo}$, a bounded observable $A_1$ of $S_1$ such
that the measured value of
$A_1$ in the state $\sigma $ determines with certainty the value of $A_2$ in
$S_2$. We determine  completely the relationship between ${\oo}$ and $\sigma
$ (Theorems 3,5 {\bf III}), and, for a fixed  state
$\sigma $ with this property, show that this predictive map 
$A_2\rightarrow A_1$ extends to a map 
$B_2\rightarrow B_1$ for all bounded observables $B_2$  lying in an
algebra canonically associated to $\sigma $, and for no others; and further
that the map that takes $B_2$ to $B_1$ is an antilinear algebra homomorphism
which is an essential imbedding (which means the kernel consists of elements
that are $0$ in the state $\sigma $ (Theorem 4, {\bf III}). Special cases of
this result have been obtained in the literature, for instance in$^{8,9}$.
Moreover, when such states exist relative to ${\oo}$, the elements of ${\oo}$
have discrete probability distributions in those states. If we now suppose, as
was done by Einstein et al, that the state
$\sigma $ has the EPR property relative to $(S_2,X_2)$ and $(S_2,P_2)$ where
$X_2,P_2$ are two bounded observables that generate the algebra of all
bounded operators (or equivalently, if the only bounded operators commuting
with both $X_2$ and
$P_2$ are the scalars), then $S_2$ has finite dimension $d\le
\dim S_1$ and the EPR states are in bijection with the set of antiunitary
isomorphisms of $S_2$ as a subsystem of $S_1$; moreover, the associated
states are essentially of the form in the example of Bohm (suitably
generalized). Of course, if we assume that the two systems have the same
finite dimension, the EPR states are completely symmetrical with respect to
the two systems, and they are exactly the generalized Bohm states (Theorem 6,
{\bf III}).
\medskip
It turns out that our definition of the EPR states forces the distributions
of the selected observables $A_2$ to be discrete. Thus the original state of
Einstein et al cannot be subsumed under our framework although it has the
same formal structure. For a rigorous discussion of this state from the
point of view of operator algebras see$^{11}$. Nevertheless one can use the
theory of approximations of quantum systems by finite quantum systems
developed in$^{13,14,15,16,17}$ to show that the Einstein--Podolsky--Rosen
state is the limit of suitably renormalized EPR states associated to a
particle moving in a large cyclic group as the order of the cyclic group goes
to infinity. For another treatment of a similar limiting process see$^{10}$.
We also mention a recent paper$^{12}$ where multipartite staes that are
maximally EPR correlated are characterized, although this appears to go
in a direction different from the line of discussion pursued in this paper.   
\bigskip\noindent {\bf II. The concept of an EPR state}  \bigskip We begin
with a brief discussion of the Bohm state and follow the discussion in pp
69---72 of$^5$. The Bohm state is that of a composite of two spin $1/2$
systems, say that of an electron and a positron, and has the form  $$ \Phi
={1\over \sqrt 2}(\varphi _+\otimes \psi _--\varphi _-\otimes
\psi _+)
$$
 $\pm $ referring to the spin up or spin down states of the electron and
positron respectively. Let $A_1$ (resp. $A_2$) denote the electron
(resp. positron) spin observable with values $\pm 1$ and corresponding
eigenstates
$\varphi _\pm
$ (resp. $\psi _\pm $). It is then a simple calculation that if in the state
$\Phi $ we know $A_1$ is observed to have a given value $\pm 1$, then the
value of $A_2$ is determined with certainty to be $\mp 1$, and vice versa.
Furthermore, let
$B_2$ be the observable in the spin system of the positron corresponding to
the spin in an arbitrary direction, so that $B_2$ has the values $\pm 1$ with
corresponding eigenstates $\eta _\pm $. Another simple calculation shows
that $\Phi $ can be expressed in the form 
$$
\Phi ={1\over \sqrt 2}(\chi _+\otimes \eta _--\chi _-\otimes \eta
_+)
$$
where $\chi _\pm $ is an orthonormal basis for the space of the electron
uniquely determined by $\eta _\pm $. Indeed, if $\eta _\pm $ are defined by
$$
\pmatrix {\psi _+\cr \psi _-\cr}=
\pmatrix {a_{11}&a_{12}\cr a_{21}&a_{22}\cr }\pmatrix 
{\eta _+\cr \eta _-\cr}
$$     
where $(a_{ij})$ is a unitary matrix, then $\chi _\pm $ are determined by
$$
\pmatrix {\chi _+\cr \chi _-\cr}=
\pmatrix {a_{22}&-a_{12}\cr -a_{21}&a_{11}\cr }\pmatrix 
{\varphi _+\cr \varphi _-\cr}
$$ 
So, if $B_2$ is the observable in the system of the positron with values
$\pm 1$ and (orthonormal) eigenstates $\eta _\pm $, then the pair of
observables $(B_1, B_2)$ has the same property as $(A_1, A_2)$, namely, that
in the state $\Phi $ if the value of $B_1$ is observed to have a given value
$\pm 1$, then the value of $B_2$ is determined with certainty to be $\mp 1$ and
vice versa. In other words, $\Phi $ has the remarkable property that if
$B_2$ is {\it any\/} observable in the positron system with values $\pm
1$, there is a uniquely associated observable $B_1$ in the electron system
such that an observation of $B_1$ that yields a value of $B_1$ predicts
the value of $B_2$ and vice versa. \medskip
The example of Bohm generalizes immediately to arbitrary finite dimensional
systems. Let ${\hh}_j (j=1,2)$ be two Hilbert spaces of the same finite
dimension $N$ and let $(\varphi _i)_{1\le i\le N}$ and $(\psi _i)_{1\le i\le
N}$ be orthonormal bases in ${\hh}_1$ and ${\hh}_2$ respectively.
${\hh}_1$ and ${\hh}_2$ are the Hilbert spaces corresponding to two
systems $S_1$ and $S_2$ respectively. Let $$ \Phi ={1\over \sqrt N}\sum _{1\le
i\le N} \varphi _i\otimes \psi _i $$
Then exactly as in the case of the Bohm example we can show that if $(\eta
_i)_{1\le i\le N}$ is any orthonormal basis of ${\hh}_2$, there is an
orthonormal basis $(\chi _i)_{1\le i\le N}$ of ${\hh}_1$ such that $\Phi $
can be expressed in the form
$$
\Phi ={1\over \sqrt N}\sum _{1\le i\le N} \chi _i\otimes \eta _i
$$
Indeed,
$$
\psi _i=\sum _ja_{ij}\eta _j\Longrightarrow \chi _i=\sum _ja_{ji}\varphi _j
$$
It
follows from this as in the Bohm example that if $B_2$ is any observable with
$N$ distinct values in the system $S_2$, there is an observable $B_1$ in the
system $S_1$ with the following property: if in the state $\Phi $ for the
compound system an observation of $B_1$ in the system $S_1$ yields an exact
value, the value of $B_2$ in $S_2$ can be predicted with certainty. It is
also remarkable  that in this and the earlier example  the roles
of
$B_1$ and
$B_2$ can be interchanged.
\medskip
\medskip
Any definition of an EPR state in the general context of two arbitrary
quantum systems will of course depend on what features of the examples of Bohm
and Einstein et al that one wishes to focus on. In order to formulate our
notion and justify its reasonableness we begin with some preliminaries.
\medskip
 Let
$S_1, S_2$ be two quantum systems and let ${\hh}_i$ be the Hilbert space of
$S_i$. As usual
${\hh}_i$ is complex and separable. Then the Hilbert space of the composite
system
$S_1\times S_2$ is the tensor product ${\hh}={\hh}_1\otimes {\hh}_2$. An
observable $A_1$ of $S_1$ is considered as an observable of $S$ via the
identification
$A_1\mapsto A_1\otimes 1$; similarly observables $A_2$ of $S_2$ are considered
as observables of $S$ via the identification $A_2\mapsto 1\otimes A_2$. Given
a state of $S_1\times S_2$, i.e., a unit vector $\sigma $ in ${\hh}_1\otimes
{\hh}_2$, the {\it commuting observables\/} $A_1\otimes 1$ and $1\otimes
A_2$ have a joint probability distribution $P^{\sigma , A_1,A_2}$
in the state $\sigma $. We shall often write $P$ or $P^\sigma $ when it is
clear what
$\sigma, A_1,A_2$ are. Then $P$ is a probability measure on
${\r}^2$; the probability measures $P_1,P_2$ induced on ${\r}$ by
the projections $(x_1,x_2)\longrightarrow x_1,x_2$ are the distributions of
$A_1\otimes 1,1\otimes A_2$ in the state $\sigma $. For borel sets $E,F$ the
probability of the event $\{A_1\otimes 1\in E,1\otimes A_2\in F\}$ is
$P(E\times F)$. We also have the family
$(q_a)_{a\in {\r}}$ of conditional probability measures on ${\r}$, with the
interpretation that
$q_a(F)$ is the conditional probability of the value of $1\otimes
A_2$ belonging to the borel set $F$ when $A_1\otimes 1$ is known to have the
value $a$. Mathematically, $(q_a)_{a\in {\r}}$ is characterized as 
the family, unique almost everywhere with respect to $P_1$, with the property
that for all borel sets $M\subset {\r}^2$,
$$
P(M)=\int _{{\r}}q_a(M[a])dP_1(a)\qquad (M[a]=\{b\
\big| \ (a,b)\in M\})
$$
\medskip
We wish to focus on the fact that the examples of Bohm and Einstein et al
feature observables $A_i$ in $S_i$ such that a measurement of $A_1$ in
$S_1$ predicts with certainty the value of $A_2$ in $S_2$. Indeed, in the
classical argumentation of Einstein et al, this property was interpreted
to mean that we can measure the observable $A_2$ in $S_2$ without
disturbing the system $S_2$. Without making this interpretation we
shall first formulate this in precise mathematical terms. Since the value
of $A_2$ is determined with certainty by the value of $A_1$ we must have a
function $g$ such that if $A_1$ is observed to have the value $a$, $A_2$
has the value $g(a)$. For general reasons we shall assume that $g$ is a
borel function. This can be formulated in either of two ways: either that 
$$
q_a(g(\{a\})=1\quad \hbox { for }P_1-\hbox {almost all }a
$$
or in the apparently weaker form where only $P$ and not the $q_a$
intervenes:
$$
P(A_1\otimes 1\in E, 1\otimes A_2\in F)=0\quad \hbox { if }
g(E)\cap F=\emptyset\ (E\cap g^{-1}(F)=\emptyset)
$$
Indeed, if the value of $A_1\otimes 1$ is $a\in E$, then the value of
$1\otimes A_2$ cannot be in $F$ if $g(E)\cap F=\emptyset$. Actually,
these two formulations are equivalent as the following lemma shows.
\bigskip\noindent
{\bf Lemma 1.\/} {\it Let $P$ be the probability measure on ${\r}^2$ as
above and let
$g$ be a borel map of
${\r}$ into ${\r}$. Let $G$ be the graph of of $g$, namely, 
$$
G=\left \{(x,g(x))\ |\ x\in {\r}\right \}
$$
Then the following statements are equivalent.
\medskip\itemitem {(a)} $P(E\times F)=0$ if $E\cap g^{-1}(F)=\emptyset$,
i.e., if $(E\times F)\cap G=\emptyset $
\smallskip\itemitem {(b)} $P({\r}^2\setminus G)=0$
\smallskip\itemitem {(c)} For $P_1$--almost all $a$, 
$$
q_a(\{g(a)\})=1
$$}
\bigskip\noindent
{\bf Proof.\/} (b)$\Longleftrightarrow$(c): It is known that $G$ is a borel
set. By general results in measure theory, $P$, which can be viewed as a
probability measure on $G$ by the condition (b), can be fibered with respect
to the projection $(x_1,x_2)\mapsto x_1$. The fibers are the points $\{g(a)\}$
and so the fiber measures are delta functions at the points $g(a)$ which is
(c). If (c) is assumed, then
$$
p(G)=\int _{{\r}}q_a(\{g(a)\})dP_1(a)=1
$$
which is (b).
\medskip
(b)$\Longleftrightarrow$(a):The implication (b)$\Longrightarrow$(a) is
trivial. The reverse implication requires a more delicate argument. However,
if $P$ is discrete, i.e., if all its mass is concentrated in a countable set,
then (a)$\Longrightarrow $(b) is easy. In fact, in this case, the probability
measures of $x_1$ and $x_2$ are both discrete. Let $D_i$ be the set of points
where $P_i$ has positive mass. Since
$P(\{a\}\times ({\r}\setminus \{g(a)\}))=0$ for $a\in D_1$ by (a), we have
$P(x_1=a,x_2=g(a))=P(x_1=a)$. Summing over $a$ one
sees that 
$P(g(D_1))=1$ and hence $P(G)=1$ which is (b). Note
that in this case
$P(x_2=g(a))\ge P(x_2=g(a), x_1=a)=P(x_1=a)>0$ so that $g$ maps $D_1$ into
$D_2$; as $P(x_2\in
g(D_1))=1$ we must have $g(D_1)=D_2$.
\medskip 
In the general case the argument for showing that (a)$\Longrightarrow $(b) is 
more technical but it is not needed for this paper (the point is that we shall
use  only the apparently weaker form (a), and as (a) is a trivial consequence
of (b) and hence also of (c), this does not affect the argumentation of the
rest of the paper). Using a general result on borel maps (see$^{6}$ p. 137)
we may assume that we are in the situation of separable metric spaces $X$ and
$Y$ and a {\it continuous\/} map $g$ of $X$ into $Y$. The probability measure
$P$ is defined on $X\times Y$ and we are given that $P(E\times F)=0$ for
borel sets $E,F$ if $(E\times F)\cap G=\emptyset $ where $G$ is the graph of
$g$. Note that the graph is now a {\it closed\/} set as $g$ is continuous
(this is also a proof that the graph of a borel map is a borel set). If
$(a,b)$ is a point not in $G$, there are open sets $E,F$ respectively
containing $a,b$ such that $E\times F$ is disjoint from $G$, and so
$P(E\times F)=0$. By separability, $X\times Y\setminus G$ can be covered by a
countable collection of sets $E_i\times F_i$ where $E_i,F_i$ are open and
$P(E_i\times F_i)=0$, and so
$P(X\times Y\setminus G)=0$. This proves that (a)$\Longrightarrow $(b).  
\bigskip\noindent
{\bf Corollary 2.\/} {\it Suppose that the equivalent conditions of the
lemma are satisfied. Then there is a borel set
$F$ such that
$F\subset g({\r})$ and $P_2(F)=1$. If $P$ is discrete, and $D_i$ is the set
of positive mass points of $P_i$, then $g(D_1)=D_2$.}
\bigskip\noindent
{\bf Proof.\/} The second statement  was established in the course of the
above proof. To prove the first note that we can find a sequence of compact
sets $G_i\subset G$ such that $P(\cup_iG_i)=1$. If $K_i$ is the image of
$G_i$ under the projection $(x_1,x_2)\mapsto x_2$, then $K_i$ is compact and
$P_2(\cup_iK_i)=1$. Obviously $\cup_iK_i\subset g({\r})$.
\medskip
We shall now make our definition of an EPR state.
\medskip
{\it Definition 1:\/} Let $A_2$ be a bounded observable of $S_2$ and 
$\sigma
\in {\hh}_1\otimes {\hh}_2$ a unit vector. Then $\sigma $ is said to be an
{\it  EPR state of $S_1\times S_2$ relative to  $(S_2, A_2)$\/} if there is 
a  bounded observable $A_1$ of $S_1$ such that $P=P^{\sigma ,A_1,A_2}$ has
the following property: there is a borel map $g({\r}\longrightarrow
{\r})$ such that  
$$
P(A_1\otimes E,1\otimes A_2\in F)=0 \quad \hbox { whenever } E\cap
g^{-1}(F)=\emptyset
$$
If there is a set ${\oo}$ of bounded observables of $S_2$ such that $\sigma
$ is EPR relative to $(S_2,A_2)$ for each $A_2\in {\oo}$, we say that $\sigma
$ is an {\it EPR state of $S_1\times S_2$ relative to $(S_2,{\oo})$.\/}  
 \bigskip\noindent 
{\bf III. The main results\/}
\bigskip
Our aim now is to explore the consequences of our definition of an EPR
state $\sigma $ relative to $(S_2,A_2)$ for the structural relationships
between
$\sigma, A_2, A_1$. Before we can formulate and prove our main results we
need some preliminaries. {\it Note that all our scalar products are linear in the first
argument and conjugate linear in the second.\/} Our entire argument   depends
on a {\it canonical\/} identification of
${\hh}_1\otimes {\hh}_2$ with the space of {\it conjugate linear\/} maps of
${\hh}_2$ into
${\hh}_1$ (equally of
${\hh}_1$ into
${\hh}_2$) that are of the Hilbert--Schmidt class. This identification is
well known, but as conjugate linear maps are somewhat less familiar than
linear ones we go into this in some detail. Let
${\cc}_{21}$ be the linear space of bounded {\it conjugate linear maps\/} 
$L({\hh}_2\rightarrow {\hh}_1)$ such that $Tr (L^{\dag } L)<\infty $. Here
$L^{\dag } $, defined by the relation $(Lu, v)=(L^{\dag } v, u)$, is also a
conjugate linear map, from ${\hh}_1$ into ${\hh}_2$, so that $L^\ast L$ is a
linear map of ${\hh}_2$. The scalar product $$ (L, M)=Tr (M^{\dag } L)\qquad
(L, M\in {\cc}_{21}) $$ then converts ${\cc}_{21}$ into a Hilbert space.
 The space ${\cc}_{21}$ contains as a
dense subspace the set ${\cc}_{21,f}$ of $L$ of finite rank 
\bigskip\noindent  {\bf Lemma 1.\/} {\it There is a canonical unitary
isomorphism  $$ \sigma \mapsto L_\sigma , \qquad {\hh}_1\otimes {\hh}_2\simeq
{\cc}_{21} 
$$ 
such that for any  $\sigma \in
{\hh}_1\otimes {\hh}_2$ and any ON basis $(e_n)$ of ${\hh}_2$,  
$$ \sigma =\sum
_nL_\sigma e_n\otimes e_n 
$$}
\bigskip\noindent 
{\bf Proof.\/}  The simplest way to
construct this canonical isomorphism is to first fix an ON basis $(e_n)$ for
${\hh}_2$. Then the elements of ${\hh}_1\otimes {\hh}_2$ are precisely those
of the form $$ \sigma =\sum _nv_n\otimes e_n\qquad (v_n\in {\hh}_1, \sum
_n||v_n||^2<\infty )\eqno (2) $$ We define $L_\sigma $ as the unique
{\it conjugate linear\/} map of Hilbert--Schmidt class of ${\hh}_2$ into
${\hh}_1$ such that
$L_\sigma e_n=v_n$. The point is that {\it $L_\sigma $ depends only on
$\sigma $ and not on the orthonormal basis $(e_n)$ that enters the
representation (2) of
$\sigma $.\/} Indeed, if $(f_m)$ is another ON basis of ${\hh}_2$, we can
write
$e_n=\sum _mu_{nm}f_m$ where $(u_{nm})$ is a unitary matrix. Then $$ \eqalign
{ \sigma &=\sum L_\sigma e_n\otimes e_n\cr 
&=\sum _{np}\overline {u_{np}}L_\sigma f_p\otimes \sum _{nm} u_{nm}f_m\cr 
&=\sum _{mp}(\sum _nu_{nm}\overline {u_{np}})L_\sigma f_p\otimes f_m\cr
&=\sum _mL_\sigma f_m\otimes f_m\cr}
$$
since
$$
\sum _nu_{nm}\overline {u_{np}}=\delta _{mp}
$$
Finally
$$
||\sigma ||^2=\sum _m||v_n||^2=\sum _n||L_\sigma e_n||^2=Tr (L_\sigma ^{\dag }
L_\sigma )\eqno (3)
$$
\bigskip\noindent
{\bf Remark 1.\/} It should be noted that had we defined $L_\sigma $ as the
{\it linear\/} map such that $L_\sigma e_n=v_n$ then it will not be
independent of the ON basis chosen. So to guarentee  the canonical nature it
is essential to choose $L_\sigma $ as the {\it conjugate linear map\/} taking
$e_n$ to $v_n$.
\bigskip\noindent
{\bf Remark 2.\/} The representation of vectors in ${\hh}_1\otimes {\hh}_2$
in the form 
$$
\sum v_n\otimes e_n\qquad ((e_n) \hbox { an ON basis of }{\hh}_2, \sum
_n||v_n||^2<\infty )
$$
is well known, see for instance the discussion of Von Neumann in Chapter VI 
of$^{12}$ where reference is made to the work of E. Schmidt. However Von
Neumann, concerned as he was about other aspects of the quantum theory of
composite systems, does not remark on the use of conjugate linear operators
that makes the representation independent of the ON basis, a fact that is
absolutely crucial for us. 
\bigskip\noindent
{\bf Remark 3.\/} The construction of the isomorphism
$$
\sigma \longmapsto L_\sigma 
$$
is perhaps not esthetically nice since we use a basis for its definition. An
alternative way is to proceed as follows. Let ${\hh}'$ be the {\it
algebraic tensor product\/} of ${\hh}_1$ and ${\hh}_2$. Then one knows that
${\hh}'$ is canonically isomorphic to the space of linear maps of finite
rank from ${\hh}_2^\ast $ to ${\hh}_1$; but ${\hh}_2^\ast $ is in canonical
{\it antiunitary isomorphism\/} with ${\hh}_2$ and so we have a canonical
linear isomorphism of ${\hh}'$ with the space of {\it conjugate linear
maps\/} of finite rank from ${\hh}_2$ to ${\hh}_1$. Explicitly, 
$$
\sigma =\sum _{1\le j\le m}a_j\otimes b_j\Longrightarrow L_\sigma u=\sum
_{1\le j\le m}(b_j,u)a_j $$
Then
$$
L_\sigma ^{\dag }w=\sum _{1\le j\le m}(a_j, w)b_j\qquad (w\in {\hh}_1)
$$
Taking the $(b_j)$ to be orthonormal, we see that $L_\sigma b_j=a_j$ and
$L_tu=0$ if $u$ is orthogonal to the $b_j$. Hence 
$$
Tr (L_\sigma ^{\dag }L_\sigma )=\sum _j||a_j||^2=||\sigma ||^2
$$
The required isomorphism is then obtained by extending the map
$\sigma \longmapsto L_\sigma $ from ${\hh}'$ to ${\cc}_{21,f}$ by completion
since ${\hh}'$ (resp. ${\cc}_{21,f}$) is dense in ${\hh}_1\otimes {\hh}_2$
(resp. ${\cc}_{21}$).
\medskip
The operator $L_\sigma ^{\dag }L_\sigma $, being of trace class, has a
discrete spectrum with eigenvalues $\lambda _j>0\ (j\ge 1)$ of finite
multiplicity, and possibly $0$ as an eigenvalue whose multiplicity could be
infinite. If ${\hh}_2(\lambda _j)$ is the eigenspace corresponding to
$\lambda _j$ and $d_j=\dim ({\hh}_2(\lambda _j))$, then
$$
Tr (L_\sigma ^{\dag }L_\sigma )=\sum _jd_j\lambda _j<\infty 
$$
We have the orthogonal decomposition
$$
{\hh}_2={\hh}_2^\sigma \oplus {\hh}_2^0
$$
where
$$
{\hh}_2^\sigma =\oplus _{j\ge 1}{\hh}_2(\lambda _j), \qquad {\hh}_2^0=\hbox {
the kernel of } L_\sigma ^{\dag }L_\sigma 
$$
We shall use these
notations a little later. At this moment we note a simple fact.
\bigskip\noindent
{\bf Lemma 2.\/} {\it Fix a unit vector $\sigma \in {\hh}_1\otimes
{\hh}_2$. Let
$B_2$ be a bounded  observable of
${\hh}_2$ commuting with $L_\sigma ^{\dag }L_\sigma $. Then, $B_2$ leaves
the ${\hh}_2(\lambda _j)$ invariant. In particular, in the state
$\sigma $ the probability distribution of $1\otimes B_2$ is discrete and is
concentrated on the set of eigenvalues of $B_2$ on ${\hh}_2^\sigma $.}
\bigskip\noindent {\bf Proof.\/} It is obvious that $B_2$ leaves the
${\hh}_2(\lambda _j)$ invariant, and as these are finite dimensional, $B_2$
has discrete spectrum on each of these and hence on ${\hh}_2^\sigma $.
Let $(e_{jp})_{1\le p\le d_j}$ be an ON basis of ${\hh}_2(\lambda _j)$
consisting of eigenstates of $B_2$, $B_2e_{jp}=b_{jp}e_{jp}$. By the previous
lemma we can write
$$
\sigma =\sum _{j\ge 1}\sum _{1\le p\le d_j}L_\sigma e_{jp}\otimes e_{jp}
$$
and so, if $\beta $ is the set of all the numbers $b_{jp}$,
$$
P^\sigma (1\otimes B_2\in \beta )\ge \sum _{j\ge 1}\sum _{1\le p\le
d_j}||L_\sigma e_{jp}||^2=Tr (L_\sigma ^{\dag }L_\sigma )=1
$$
\medskip
We now come to the result which is the basis for everything that we can
say about EPR states. Its proof depends
essentially on the possibility of using {\it any\/} ON basis of ${\hh}_2$ in
the decomposition of $\sigma $. 
\bigskip\noindent
{\bf Theorem 3.\/} {\it Let ${\oo}$ be any set of bounded observables
of ${\hh}_2$ and let $\sigma
$ be an element of unit norm in
${\hh}_1\otimes {\hh}_2$. If 
$L_\sigma $  is the element of ${\cc}_{21}$ that corresponds to $\sigma $
under the canonical isomorphism of Lemma 1, then $\sigma $ is an EPR state
relative to $(S_2,{\oo})$ if and only if 
$L_\sigma ^{\dag } L_\sigma $ commutes with every element of ${\oo}$.}
\bigskip\noindent
{\bf Proof.\/}  It is obviously enough to do this for each element of
${\oo}$ separately. Fix $B_2\in {\oo}$ and assume that $\sigma $ is EPR
relative to $(S_2,B_2)$. Let
$B_1$ be a bounded observable of ${\hh}_1$ with the following property: there
is a borel map $g$ of ${\r}$ into ${\r}$ such that 
$$
P(B_1\otimes 1\in E,1\otimes B_2\in F)=0\qquad (E\cap
g^{-1}(F)=\emptyset )
$$
We should prove that $L_\sigma ^{\dag } L_\sigma $ commutes with $B_2$.  The
proof is slightly simpler if
$B_1$ and
$B_2$ have discrete spectra, but not  by much. Still it may be worthwhile to
give the argument separately in this case.
\medskip
{\it Case of discrete spectra:\/} Let $\beta _1 $ (resp. $\beta _2$) be the
set of eigenvalues of $B_1$ (resp. $B_2$). For $a\in \beta _1$ (resp. $b\in
\beta _2$) let $E_a$ (resp. $F_b$) be the corresponding eigenspace. Then $g$
is
 a map $\beta _1\longrightarrow \beta _2$. 
\medskip
We are given that
$$
P(B_1\otimes 1=a,1\otimes B_2=b)=0 \qquad (b\not=g(a))
$$
Fix $b\in \beta _2
$. Select an ON basis $(e_i)$ of $F_b$ and an ON basis
$(f_j)$ of $F_b^\perp$ and write
$$
\sigma =\sum _iL_\sigma e_i\otimes e_i+\sum _jL_\sigma f_j\otimes f_j
$$
Let $Q_a$ be the orthogonal projection ${\hh}_1\longrightarrow E_a$. Then
$$
\eqalign { P(B_1\otimes 1=a,1\otimes B_2=b)&=||\sum
_iQ_aL_\sigma e_i\otimes e_i||^2\cr
&=\sum _i||Q_aL_\sigma e_i||^2\cr}
$$
Since this is zero for $b\not=g(a)$, we must have
$$
Q_aL_\sigma e_i=0\qquad (b\not=g(a))
$$
In other words, if we write
$$
E[b]=\oplus _{a : g(a)=b}E_a
$$
then
$$
L_\sigma [F_b]\subset E[b]
$$
Suppose now that $b'\not=b$, and let $u\in F_b, v\in F_{b'}$. Then
$$
(L_\sigma ^{\dag}L_\sigma u,v)=(L_\sigma v, L_\sigma u)=0
$$
since
$$
E[b]\perp E[b']
$$
Thus
$$
L_\sigma ^{\dag}L_\sigma u\in F_b
$$
This proves that $L_\sigma ^{\dag}L_\sigma $ leaves all the $F_b$ invariant
and hence that it commutes with $B_2$.
\medskip
{\it General case :\/} We must prove that $L_\sigma ^{\dag}L_\sigma $
commutes with all the spectral projections of $B_2$. Since $L_\sigma
^{\dag}L_\sigma $ is self adjoint, this is equivalent to showing that
$L_\sigma ^{\dag}L_\sigma $ leaves the spectral subspaces of $B_2$ invariant.
For any borel set $B\subset {\r}$ let $F_B$ (resp. $E_B$) be the corresponding
spectral subspace of $B_2$ (resp. $B_1$). Write $Q_B$ for the orthogonal
projection ${\hh}_1\longrightarrow E_B$. Fix a borel set
$B\subset {\r}$. Select ON bases $(e_i)$ for $F_B$ and $(f_j)$ for
$F_B^\perp$. Then
$$
\sigma =\sum _iL_\sigma e_i\otimes e_i+\sum _jL_\sigma f_j\otimes f_j
$$
If $C=g^{-1}(B)$, then
$$
\eqalign { 
0&=P(B_1\otimes 1\in {\r}\setminus C,1\otimes B_2\in B)\cr
&=||\sum _iQ_{{\r}\setminus C}L_\sigma e_i\otimes e_i||^2\cr
&=\sum _i||Q_{{\r}\setminus C}L_\sigma e_i||^2\cr}
$$
and hence
$$
Q_{{\r}\setminus C}L_\sigma e_i=0\qquad (\hbox { for all } i)
$$
Thus
$$
L_\sigma [F_B]\subset E_{g^{-1}(B)}
$$
We now calculate $(L_\sigma ^{\dag}L_\sigma u,v)$ for $u\in F_B, v\in
F_{{\r}\setminus B}$. We have
$$
(L_\sigma ^{\dag}L_\sigma u,v)=(L_\sigma v, L_\sigma u)=0
$$
since $g^{-1}({\r}\setminus B)={\r}\setminus g^{-1}(B)$ and
$$
E_{g^{-1}(B)}\perp E_{{\r}\setminus g^{-1}(B)}
$$
Thus
$$
L_\sigma ^{\dag}L_\sigma [F_B]\perp F_{g^{-1}(B)}
$$
which gives
$$
L_\sigma ^{\dag}L_\sigma [F_B]\subset F_B
$$
This is what we wanted to prove.
\medskip
We now take up the converse. We assume that $B_2$ commutes with $L_\sigma
^{\dag } L_\sigma $ and wish to find a bounded observable $B_1$ of ${\hh}_1$
such that the EPR property is satisfied for the pair $(B_1,B_2)$. We use
Lemma 2 above.  On ${\hh}_2(\lambda _j)$ we can write $L_\sigma $ as $\lambda
_j^{1/2}U_j$ where $U_j$ is an {\it antiunitary imbedding\/} of
${\hh}_2(\lambda _j)$ into
${\hh}_1$. If ${\hh}_1(\lambda _j)=L_\sigma [{\hh}_2(\lambda _j)]$, it is
then easy to check that the ${\hh}_1(\lambda _j)$ are mutually orthogonal. Let
$$
{\hh}_1^\sigma =\oplus _j{\hh}_1(\lambda _j)
$$
We define $U$ as the antiunitary isomorphism of
${\hh}_2^\sigma $ with ${\hh}_1^\sigma \subset {\hh}_1$ which is equal to
$U_j$ on
${\hh}_2(\lambda _j)$. If now 
$(e_{jp})_{1\le p\le d_j}$ is any ON basis of ${\hh}_2(\lambda _j)$, we have
the representation
$$
\sigma =\sum _{j\ge 1}\lambda _j^{1/2}\sum _{1\le p\le d_j}Ue_{jp}\otimes
e_{jp}
$$
We take the $e_{jp}$ to be the eigenstates of $B_2$,
$B_2e_{jp}=b_{jp}e_{jp}$. Let us  define 
$$
B_1=UB_2U^{\dag}
$$
It is easy to check that $B_1Ue_{jp}=b_{jp}Ue_{jp}$. We may therefore conclude
that the distribution of $B_1\otimes 1$ is discrete in the state
$\sigma $ with its mass concentrated on the set $\beta _2$ of eigenvalues of
$B_2$ in
${\hh}_2^\sigma $. It is  immediate that 
$$
P(B_1\otimes 1=b_1,1\otimes B_2=b_2)=0\qquad (b_1\not=b_2, b_i\in \beta _2)
$$
This completes the proof of the theorem.
\bigskip\noindent
{\bf Remark.\/} Note the obvious symmetry between the roles of $B_1$ and
$B_2$ as revealed in the last realtion.
\medskip
For any set ${\oo}$ of bounded observables in ${\hh}_2$ we write ${\oo}'$
for the set of bounded observables commuting with ${\oo}$ and
${\oo}''=({\oo}')'$. Theorem 3 leads at once to the following results.  
\bigskip\noindent
{\bf Theorem 4.\/} {\it Let notation be
as above and let
$\sigma
$ be a unit vector in ${\hh}_1\otimes {\hh}_2$. Let $L_\sigma \in {\cc}_{21}$
correspond to $\sigma $ and let ${\bb}^\sigma $ be the Von Neumann algebra of
bounded operators commuting with $L_\sigma ^{\dag}L_\sigma $. Then $\sigma $
is EPR relative to the observables in ${\bb}^\sigma $ and to no others. These
observables all have discrete probability distributions in the state $\sigma
$, which are  concentrated on the set of their eigenvalues in ${\hh}_2^\sigma
$ (on which they have discrete spectra). The state $\sigma $ induces an
antilinear homomorphism $B_2\mapsto B_1$ of ${\bb}^\sigma $ into the algebra
of bounded operators of ${\hh}_1$, and the distributions of $B_1$ and $B_2$
are the same for all observables in $B_2\in {\bb}^\sigma $. Moreover we have
$$
P(B_1\otimes =b_1,1\otimes B_2=b_2)=0\qquad (b_1\not=b_2,\ b_i\in \beta
^\sigma )
$$
where $\beta ^\sigma $ is the spectrum of $B_2$ on ${\hh}_2^\sigma $.}
\bigskip\noindent
{\bf Remark.\/} The map $B_2\mapsto B_1$ need not be an imbedding. However
all observables in its kernel vanish on ${\hh}_2^\sigma $ and so vanish with
probability $1$ in the state $\sigma $. We may therefore say that it is an
{\it essential imbedding.\/}
\medskip
Intuitively, the existence of the essential imbedding of ${\bb}^\sigma $
inside $S_1$ is reasonable because, as $B_1$ determines $B_2$, the
propositions of $B_2$ must be found within those of $B_1$, and so, by
Wigner's theorem, this map should be effected by a symmetry. The technical
point  which goes beyond this heuristic reasoning is that this symmetry is
antiunitary.
\bigskip\noindent
{\bf Theorem 5.\/} {\it If ${\oo}$ is any set of bounded observables of
${\hh}_2$, there exist EPR states relative to $(S_2,{\oo})$ if and only if
there are projections $Q$ commuting with ${\oo}$ whose ranges have dimensions
$\le \dim {\hh}_1$.   If $Q_j$ is a family of such projections which are
mutually orthogonal,
$F_j=\hbox { range of } Q_j$, $d_j=\dim F_j$, and if $\dim \oplus F_j\le \dim
{\hh}_1$, then for any set of numbers $d_j$ such that $\sum _jd_j\lambda
_j=1$ and any antiunitary imbedding $U$ of $\oplus F_j$ into ${\hh}_1$ the
state 
$$
\sigma =\sum _{j\ge 1}\lambda _j^{1/2}\sum _{1\le p\le d_j}Ue_{jp}\otimes
e_{jp}
$$
where the $e_{jp}$ are any ON basis for ${\hh}_2(\lambda _j)$ is EPR relative
to $(S_2,{\oo})$. Every state EPR relative to $(S_2,{\oo})$ is obtained this
way, and any such is EPR relative to $(S_2, [{\oo}])$ where $[{\oo}]$ is the
set of observables in the Von Neumann algebra generated by ${\oo}$.} 
\bigskip\noindent
{\bf Remark.\/} The fact that a state which has the EPR property with respect
to some observables has that property for infinitely many others has been
known for a long time; see$^{8,9}$.
\medskip
Suppose we assume, as is the case in the Einstein--Podolsky--Rosen
example, that the bounded observables $X_2$ and $P_2$ of ${\hh}_2$ have the
property that the only bounded observables simultaneously measurable with
both of them are the scalars. Then $L_\sigma ^{\dag}L_\sigma $ must be a
scalar, there is only one
$j$ in the above formulae, ${\hh}_2={\hh}_2(\lambda _1)$ is finite
dimensional, and $\dim {\hh}_1\ge \dim {\hh}_2$. Then
$U$ is an antiunitary injection of ${\hh}_2$ into ${\hh}_1$ and there is a
bijection between EPR states of $S$ relative to $(S_2, \{X_2,P_2\})$ and the
equivalence classes of antiunitary imbeddings of ${\hh}_2$ into ${\hh}_1$. In
particular, if, as is true in many examples, that $S_1$ and $S_2$ are
identical, then the EPR states are the same relative to each system, and are
in canonical bijection with the set of antiunitary symmetries between the
two systems. We thus have the following theorem.
\bigskip\noindent
{\bf Theorem 6.\/} {\it Let $\sigma $ be a unit vector in ${\hh}_1\otimes
{\hh}_2$ and let it be an EPR state relative to $(S_2,\{X_2,P_2\})$ where
$X_2,P_2$ are bounded observables with the property that only the scalars
in $S_2$ commute with both of them. Then $\dim {\hh}=d<\infty $, $\dim
{\hh}_1\ge d$, and there is an antiunitary imbedding $U$ of ${\hh}_1$ into
${\hh}_2$ such that
$$
\sigma =d^{-{1/2}}\sum _{1\le j\le d}Ue_j\otimes e_j
$$
where $(e_j)$ is any ON basis of ${\hh}_2$. The correspondence $\sigma
\rightarrow U$ induces a bijection between the set of states of $S_1\times
S_2$ that are EPR relative to $(S_2,X_2,P_2)$ and the set of equivalence
classes of antiunitary imbeddings of ${\hh}_2$ into ${\hh}_1$. In this case,
if
${\bb}$ is the set of all bounded observables of $S_2$ and, for $B_2\in {\bb}$
we define $B_1=UBU^{\dag}$, then $B_1\otimes 1$ and $1\otimes B_2$ have
identical distributions in $\sigma $ which are concentrated on the
(finite) set $\alpha $ of eigenvalues of $B_2$, and
$$
P(B_1\otimes 1=b_1,1\otimes B_2=b_2)=0\qquad (b_1\not=b_2, b_1,b_2\in
\alpha )
$$
If further $\dim {\hh}_1=\dim {\hh}_2$, the EPR states  relative
to $(S_2,\{X_2,P_2\})$ are precisely those that are EPR relative to
$(S_1,X_1,P_1)$, and are in bijective correspondence with the set of
antiunitary isomorphisms of ${\hh}_1$ and ${\hh}_2$.}
\bigskip\noindent
{\bf Remark.\/} We see that in this case the EPR states are essentially
the same as the ones discussed at the beginning of this paper as
generalizations of the Bohm states.     
\bigskip\noindent {\bf IV. Examples\/}
\bigskip
We have mentioned already that one can construct the  analogues of the
original Einstein--Podolsky--Rosen states in certain finite quantum systems.
These finite systems  were  first introduced by  Weyl$^{13}$ and explored
subsequently by Schwinger$^{14}$, Digernes et al$^{15}$,  Husstad$^{16}$,
and Digernes et al$^{17}$. In their most
general form they treat a particle which moves not in the real line ${\r}$
but in a {\it finite abelian group\/} $G$ (for Weyl and Schwinger this group
was
${\z}_N$, the group of integers modulo $N$ while for Digernes et al it was a
more general finite abelian group). When $G$ is a large cyclic group it
serves as an approximation to ${\r}$, which is the point of view of the
papers loc. cit. Indeed, the group ${\z}_N$ is identified by a grid of
equidistant points symmetric about the origin in ${\r}$, with the intergrid
distance of the order of $N^{-1/2}$, so that when $N\to \infty $ the
kinematics of the system go over in the limit to the kinematics of  the usual
one particle system in quantum mechanics. We shall take up this approximation
point of view in the next section. Here we shall keep our  discussion to the
structure of some specific EPR states. We take
${\hh}_1={\hh}_2=L^2(G)$ where $G$ is a finite abelian group whose order will
be denoted by $|G|$. The scalar product is given by $$ (f,g)={1\over |G|} \sum
_{x\in G} f(x)g(x)^{\rm conj}\qquad (f,g\in L^2(G)) $$ For simplicity we
consider the antiunitary isomorphism $$ U: f\longmapsto f^{\rm conj} $$   of
${\hh}_2$ with ${\hh}_1$. If $(e_n)$ is any ON basis of ${\hh}_1$ we have the
representation of the corresponding state $\sigma _U$ as $$ \sigma
_U=d^{-1/2}\sum e_n^{\rm conj}\otimes e_n  $$
Now we have two ON bases of ${\hh}_2$, namely
$$
\{\sqrt {|G|}\delta _x\}_{x\in G}, \qquad \{\xi \}_{\xi \in \widehat G}
$$
where $\delta _x$ is the delta function at $x$ and $\widehat G$ is the
group of characters of $G$. So
$$
\sigma _U=\sqrt {|G|} \sum _{x\in G}\delta _x\otimes \delta _x={1\over
\sqrt {|G|}}\sum _{\xi \in \widehat G}\xi ^{-1}\otimes \xi \eqno (\ast )
$$
The equality of the last two expressions in ($\ast $) can also be verified
directly using the orthogonality relations in $G$ and $\widehat G$. Let $X_2$
be an observable in $S_2$ with distinct values $a_x(x\in G)$ and corresponding
eigenstates $\delta _x$. Then $UX_2U^{\dag } =X_1$ has the same definition in
${\hh}_1$. Clearly $X_i$ are the {\it position\/} observables in the two
systems. For
$Y_2$ we take the observable in $S_2$ with distinct values $b_\xi $ and
eigenstates
$\xi $. Then $Y_1=UY_2U^{\dag }$ is the observable in ${\hh}_1$ with values
$b_\xi $ and eigenstates $\xi ^{-1}$. The $Y_i$ are the momentum observables
in the two systems. It is then a simple calculation to verify the EPR
property. For the pair $({\hh}_1, {\hh}_2)$ these are summarized by the
symmetrical relations
$$
\eqalign {P(X_1\otimes 1=a_x, 1\otimes X_2=a'_x)&=0\quad (a_x\not=a'_x)\cr
 P(Y_1\otimes 1=b_\xi,1\otimes Y_2=b'_\xi )&=0\quad (b_\xi
\not=b'_\xi )\cr} 
$$
\medskip 
As a second example let us take ${\hh}_1={\hh}_2=L^2(G)\otimes {\c}^N$ and
let the bounded observables $X_2',P_2'$ of ${\hh}_2$ be defined by
$X_2'=X_2\otimes 1,P_2'=P_2\otimes 1$, the observables $X_2,P_2$ in $L^2(G)$
being as in the preceding example (all this inside ${\hh}_2$, the
tensor products here should not be confused with the one involving ${\hh}_1$
and ${\hh}_2$). One may view this as a quantum system of a
particle with $N$ spin states moving in the finite abelian group $G$. The
commutator of $\{X_2',P_2'\}$ is then the algebra ${\ab}=1\otimes {\mm}$ where
${\mm}$ is the matrix algebra in ${\c}^N$. Although $X_2'$ and $P_2'$
generate an algebra without any dispersion states, nevertheless there is a
wide choice of EPR states relative to $\{X_2',P_2'\}$ since the choice of
$L_\sigma ^{\dag}L_\sigma $ within 
${\ab}$ is arbitrary, so that they will depend on more than just an
antiunitary imbedding of ${\hh}_2^\sigma $ into ${\hh}_1$. This example shows
that the structure of EPR states relative to a set ${\oo}$ does not depend 
exclusively on the structure of the algebra generated by ${\oo}$ but also on
its commutator in
${\hh}_2$.
\bigskip\noindent
{\bf V. States of infinite norm\/}
\bigskip
Theorem 5 of {\bf IV} shows that if an observable $B_2$ in ${\hh}_2$
has a continuous spectrum, there is no EPR state relative to it. Strictly
speaking therefore, the original state  of Einstein et al is not
subsumed under our results since their state is defined by a tempered 
distribution which does not have finite norm. Nevertheless 
in the approximation scheme of Weyl, Schwinger and others mentioned in the
previous section, in the limit when
$G={\z}_N$ approaches ${\r}$, the
states considered there might be expected to  go over to the original
Einstein--Podolsky--Rosen state {\it after a renormalization\/}. We shall see 
now that this is the case. Since  the calculations are similar to those found
in$^{14,15}$ we shall be very brief. Indeed, in this example,
${\hh}_1={\hh}_2=L^2({\r})$ and the identification is taken to be, as in the
finite case, the map  $$ 
U : \psi \longmapsto \psi ^{\rm
conj}\qquad (\psi \in {\hh}_2) 
$$ 
Let $X_i$ (resp. $P_i$) be the position (resp. momentum) of the $i^{\rm th}$ 
particle. The original state of Einstein et al is
$$
\sigma _U=\sigma =\int e^{i(x_1-x_2)p}dp\qquad (\hbar =1) 
$$
which can be written as 
$$
\int Ue_p\otimes e_pdp,\qquad e_p(x)=e^{-ixp}
$$
Of course the integrals have to be interprted as tempered
distributions and so have to be paired with Schwartz functions. In this
state, if $P_1\otimes 1=p$, then $1\otimes P_2=-p$.  A simple calculation
using Fourier analysis then shows that we also have the representation
$$
\sigma =\int \delta _x\otimes \delta
_x dx
$$
Thus
$$
\sigma =\int Ue_p\otimes e_pdp=2\pi \int \delta _x\otimes
\delta _x dx
$$
The analogy with ($\ast $) of section {\bf IV} is now clear. This
representation shows that in this state, if $X_1\otimes 1=x$, then $1\otimes
X_2=x$. 
\medskip
To exhibit $\sigma $ as the limit of renormalized EPR states associated to
the cyclic group ${\z}_N$ we use the imbedding of $L^2({\z}_N)$ into
$L^2({\r})$ given by (see$^{15})$
$$
N^{1/2}\delta _x\longmapsto \varepsilon ^{-1/2}\chi _{r\varepsilon }
$$
where $\varepsilon =(2\pi /N)^{1/2}$, and $\chi _{r\varepsilon }$ is the
characteristic function of the interval $((r-1/2)\varepsilon ,
(r+1/2)\varepsilon )$ and $x$ runs through the congruences classes of
$r(r=0,\pm 1,\pm 2,\dots ,\pm (N-1)/2)$ (we take $N$ to be odd, which is of
no consequence as we let $N\to
\infty $). Then the EPR state associated to ${\z}_N$ is
$$
N^{1/2}\sum _x\delta _x\otimes \delta _x
$$
which goes over under our imbedding to
$$
\sigma _N=(2\pi )^{-1/2}\sum _{|r|\le (N-1)/2}\chi _{r\varepsilon }\otimes
\chi _{r\varepsilon }
$$
Since, for any Schwartz function $f$ we have
$$
\langle \chi _{r\varepsilon }, f\rangle =\int _{((r-1/2)\varepsilon
}^{(r+1/2)\varepsilon }f(x)dx=\varepsilon f(r\varepsilon )+O(\varepsilon
^3)
$$
we find that
$$
\eqalign {
N^{1/2}\sigma _N(f\otimes g) &=\sum _{|r|\le
(N-1)/2}\varepsilon f(r\varepsilon )g(r\varepsilon )+O(\varepsilon)\cr
&\longrightarrow \int _{{\r}}f(x)g(x)dx\cr}
$$
Hence
$$
N^{1/2}\sigma _N\longrightarrow \sigma \qquad (N\to \infty )
$$
Note that the norm of the state on the left goes to infinity as it should,
since the left side is a state of infinite norm.
\medskip
As we mentioned in the introduction, the
paper$^{10}$ contains a detailed discussion of the original EPR state of
Einstein et al as a limit of normalized states with very sharp correlations
between the position and momentum variables in the two systems, while the
paper$^{11}$ contains a rigorous characterization of the
Einstein--Podolsky--Rosen state from the point of algebraic quantum theory.
\medskip
It is  easy to see that we can generalize the original example of Einstein
et al by taking other choices of
$U$. If
$U$ is
 an antiunitary isomorphism of the Schwartz space with itself
in the Schwartz topology then we obtain a class of states
generalizing the example of Einstein et al. For instance we may
take
$$
(Uf)(x)=e^{i\theta (x)}f(x)^{\rm conj}
$$
where $\theta $ is a smooth real function whose derivatives have
polynomial growth at  most. We shall take up the properties of these states
on a later occasion.
\bigskip\noindent {\bf Acknowledgments\/}
\bigskip
The authors are grateful to Professors E. G. Beltrametti and Professor G.
Cassinelli of the Departimento di Fysica of the University of Genova and the
Istituto Nazionale di Fysica Nucleare in Genova for  their reading of an
earlier version of this paper and their valuable comments. The authors also
wish to express their deep gratitude to the referee. The referee  supplied us
with many important references to the literature where related questions had
been treated, and his insistence that the mathematics should correspond as
closely as possible to the physics led to major improvements in both
the style and substance of this paper.  
\vskip 0.7 true in 
{\mysmall \item
{$^1$}Albert Einstein, Boris Podolsky, and Nathan Rosen, Phys. Rev.,{\bf 47},
777 (1935). \smallskip   \item {$^2$}Niels Bohr, Phys.
Rev.,{\bf 48}, 696 (1935).
\smallskip 
\item {$^3$}J. A. Wheeler and W. H. Zurek, {\eightit
Quantum Theory and Measurement\/}, Princeton University Press , 1983. 
\smallskip 
\item {$^4$}D. Bohm, {\eightit Quantum Theory,\/}, Prentice--Hall (1951).
\smallskip 
\item {$^5$}E. G. Beltrametti, and G. Cassinelli, {\eightit The Logic of
Quantum Mechanics,\/} Ch. 7, Encyclopedia of Mathematics and its Applications, Vol
15, Addison--Wesley, 1981.
\smallskip 
\item {$^6$}K. R. Parthasarathy, {\eightit Probability measures on Metric
Spaces\/}, Academic Press, 1967.
\smallskip 
\item {$^7$}J. Von Neumann, {\eightit Mathematical Foundations of Quantum
Mechanics\/}, Princeton University Press, 1955.
\smallskip 
\item {$^8$}E. Schr\"odinger, Proc. Camb. Phil. Soc., {\bf 31} (1935), 555;
Proc. Camb. Phi. Soc., {\bf 32}, 446.
\smallskip 
\item {$^9$}B. C. Van Fraassen, {\eightit Quantum Mechanics,\/} Oxford, 1991.
\smallskip 
\item {$^{10}$}O. Cohen, Phys. Rev. A, {\bf 56} (1997), 3484. 
\smallskip 
\item {$^{11}$}H. Halvorson, and R. Clifton, lanl archives, quant-ph/9905042.
\smallskip 
\item {$^{12}$}R. Clifton et al, Phys. Rev. A, {\bf 58} (1998),135.
\smallskip  
\item {$^{13}$}H. Weyl, {\eightit Theory of Groups and Quantum Mechanics\/},
Dover, 1931, Ch. III, \S 16, Ch. IV, \S\S14, 15. \smallskip 
\item {$^{14}$}J Schwinger,
{\eightit Quantum Kinematics and Dynamics,\/} W. A. Benjamin, New York, 1970. 
\smallskip 
\item {$^{15}$}T. Digernes, V. S. Varadarajan, and S. R. S.
Varadhan,  Rev. Math.
Phys., {\bf 6}, 621 (1994).
\smallskip\item {$^{16}$}E. Husstad, {\eightit Endeligdimensjonale
approksimasjoner til kvantesystemer\/}, Thesis, University of Trondheim,
1991/92. \smallskip
\item {$^{17}$}T. Digernes, E. Husstad, and V. S. Varadarajan (preprint),
Math. Scand. (to appear).
\medskip}

\bye